\documentstyle[twoside,fleqn,npb,epsfig,feynmp,amssymb]{article}
\newcommand{\eq}{\begin{equation}} 
\newcommand{\eqx}{\end{equation}}
\newcommand{\eqn}{\begin{eqnarray}} 
\newcommand{\eqnx}{\end{eqnarray}}

\newcommand{\f}[2]{\frac{#1}{#2}}

\newcommand{\cor}[1]{\left\langle{#1}\right\rangle}

\newcommand{\dl}{\delta}

\newcommand{\al}{\alpha}
\newcommand{\bt}{\beta}

\newcommand{\CC}{\mathbb{C}}
\newcommand{\ZZ}{\mathbb{Z}}

\newcommand{\rrr}{\mathbb{R}}

\newcommand{\EE}{{\cal E}}

\newcommand{\slc}{$SL(2,\CC)$\ }

\newcommand{\ehk}{\EE^h_{k,q}}

\newcommand{\ket}[1]{\left|{#1}\right\rangle}
\newcommand{\vac}{\ket{0}}

\newcommand{\raz}{\rho_{a_0}}
\newcommand{\rao}{\rho_{a_1}}
\newcommand{\rbz}{\rho_{b_0}}
\newcommand{\rbo}{\rho_{b_1}}
\newcommand{\rpz}{\rho_{p_0}}
\newcommand{\rpo}{\rho_{p_1}}

\newcommand{\rr}[4]{#1, {\it #2 \/}{\bf #3} #4}

\newcommand{\AmS}{{\protect\the\textfont2
  A\kern-.1667em\lower.5ex\hbox{M}\kern-.125emS}}

\title{Hard diffraction and QCD multi-Pomeron vertices}
\author{R. Peschanski\address{CEA, Service de Physique Th\'eorique,
CE-Saclay\\
F-91191 Gif-sur-Yvette Cedex, France}
}

\begin{document}
\begin{abstract}
We discuss the phenomenological and theoretical implications of recent progresses in the evaluation of multi-Pomeron vertices in high-energy perturbative QCD.
\end{abstract}
\maketitle
\section{Multi-Pomeron vertices in the QCD dipole model}
The QCD dipole model \cite {mueller} happens to be a quite useful representation
of the QCD perturbative resummation at leading logarithms at high incident energy
(or, equivalently, at small Bjorken $x$) known as the BFKL QCD Pomeron \cite{bfkl}. On a phenomenological ground, it provides an ``s-channel'' description of hard diffraction \cite {bp} which appears to be quite successful in describing the present data at HERA \cite {munier}. On a theoretical ground, the dipole model represents the $1/N_c$ limit of perturbative QCD, and leads to  interesting simplifications in the calculation of triple \cite {value} and multiple \cite {multiple} Pomeron vertices.  The present contribution gives an overview of these phenomenological and theoretical results. 
\section{Theory}
The main result of applying the QCD dipole model to the calculation of 
multi-Pomeron vertices can be briefly described as follows \cite {multiple}:
The $1\!\to\!p$ Pomeron vertex can be obtained \cite {pesch} from the calculation of the  QCD dipole multiplicity
density (i.e. the probability for finding $p$ dipoles $\rbz \rbo,$...$,\rpz \rpo$ in an initial one $\raz \rao$) coming from the solution of a integro-differential equation. The solution reads:
\eqn
B_{1\to p}=\int \frac{d^{2}\rho _{0}...d^{2}\rho _{p}}
{\left| \rho _{01}\ \rho _{12} ... \rho _{p0}\right| ^{2}}\times\nonumber\\ E^{h_0}\left( \rho _{0\alpha},\rho _{1\alpha} \right)... E^{h_p}\left( \rho _{p\pi},\rho _{0\pi}\right), 
\label{2}
\eqnx
with $\rho _{ij}=\rho _{i}\!-\!\rho _{j} \left({\rm  resp.}\ \bar{\rho}
_{ij}=\bar{\rho _{i}}\!-\!\bar{\rho _{j}}\right) .$ The $\rho_{\alpha}...\rho_{\pi}$ are auxiliary variables which play the r\^ole of dipole c.o.m. coordinates. The $E^{h}\left( \rho _{i\delta },\rho _{j\delta }\right) =
\left( -1\right)^{n}
\left( \frac{\rho _{ij}}{\rho _{i\delta }\rho _{j\delta }}\right)^{h}
\times \left( \frac{\bar{\rho} _{ij}}{\bar{\rho} _{i\delta }\bar\rho _{j\delta}}\right)^{\tilde{h}},
$  are \cite{Lipatov} the \slc eigenvectors
labeled by the quantum numbers of the irreducible unitary representations, namely $h =i\nu +\frac {1-n}2$, $ \tilde h
= 1\!- \!\bar h = i\nu +\frac {1+n}2,$ ($n\!\in \! {\ZZ}$, ${\nu }\!\in  \!{\rrr}$). Interestingly enough,
 the expression of   the functions $B_{1\to p}$  can be considered as  correlation functions, namely 
\eq
\label{e.dipcor}
B_{1\to p} \equiv \cor{0|\Phi^{h_0}(\rho_\al) \Phi^{h_1}(\rho_\bt) ...\Phi^{h_p}(\rho_\pi)|0}
\eqx
where the $\Phi^h(\rho)$ are   suitably defined operators. An explicit construction following e.g. the BPZ construction \cite{bpz} can be given \cite {multiple}, as briefly summarized now.

A substantial simplification occurs when we  consider the Fourier
transform of the $\Phi^h$ operator in momentum space:
\eq
\Phi^h(q)=\int d^2x \ e^{-iqx}\Phi^h(x)\ .
\eqx
One may define
the Hilbert space, vaccuum state and operators in such a way that:
\eqn
\cor{\Phi^{h_0}(q_0)\ldots\Phi^{h_p}(q_p)}= \nonumber\\
\dl(q_0+\ldots+q_p) \cor{\phi^{h_0}(q_0)\ldots \phi^{h_p}(q_p)},\label{corfourier}
\eqnx
where the $\phi^h(q)$ act on the Hilbert space $ f \in L^2(\CC)$ with the
vacuum $\vac=1$ by
\eq
\label{e.constrfin}
[\phi^h(q) f](k)= \ehk \cdot f(k-q) 
\eqx
and
\eq
\ehk\equiv \int d^2\rho d^2\rho' \ e^{iq\rho'+ik(\rho\!-\!\rho')}\ 
\f{E^h(\rho,\rho')}{|\rho\!-\!\rho'|^2} 
\label{Epsilon}
\eqx
can be explicitely calculated in terms of hypergeometric functions \cite{multiple}.
The  correlation functions read
\eqn
\label{e.aform}
\cor{\phi^{h_0}(q_0) ...
\phi^{h_p}(q_p)} = 
\int d^2k \ \EE^{h_0}_{k,q_0} \times\nonumber\\
\EE^{h_1}_{k-q_0,q_1}... \EE^{h_p}_{k-q_0-...-q_{p-1},q_p}\ .
\eqnx
All dipole correlators and thus QCD Pomeron vertices are just expressed by a
{\em single} integral over a product of  $\ehk$ functions. 

The expression (\ref{e.aform}) allows a rather simple and attractive representation of Pomeron vertices (see Fig.1) in terms of a one-loop integral in momentum space with vertices defined by the functions $\ehk.$ Each Pomeron    interacts in momentum space {\it via} a 3-vertex defined by a function $\ehk$ or $\bar \ehk$ depending whether it is 
created or annihilated. The  momentum is conserved at each vertex.

\begin{figure}[hbt]
\vspace{9pt}
\framebox[75mm]{\rule[0mm]{0mm}{75mm}

\epsfig{file=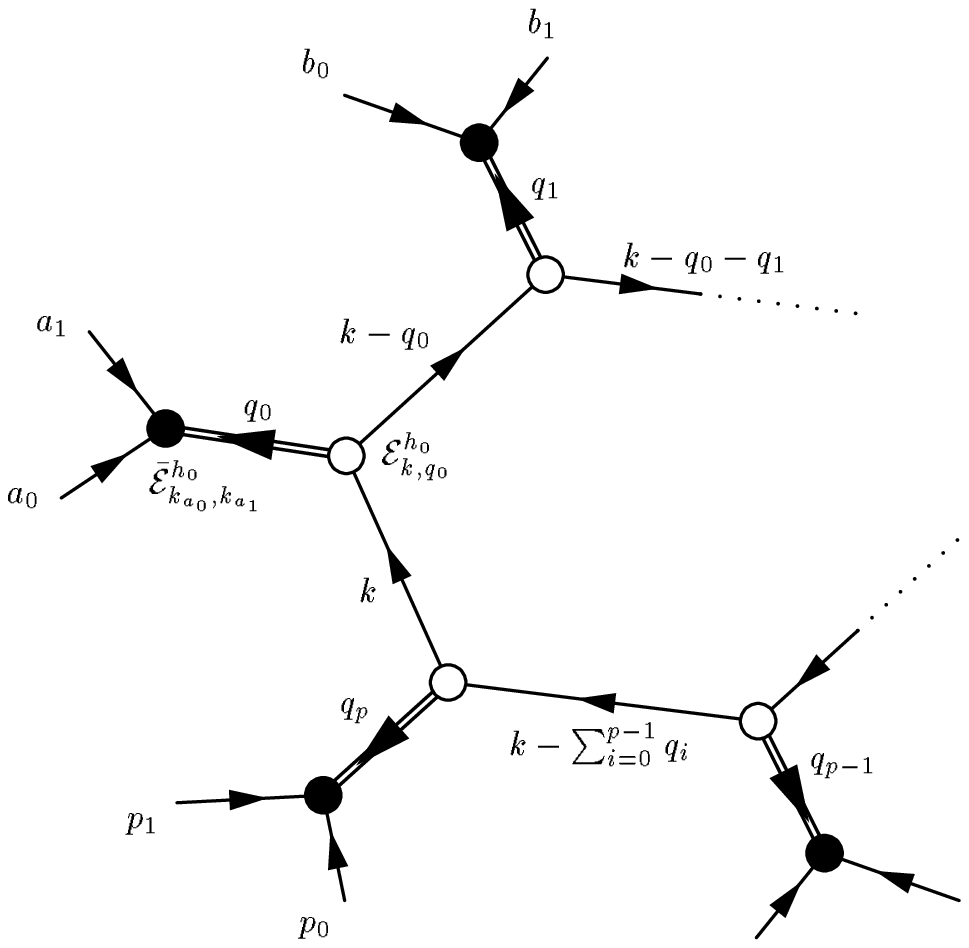,height=70mm}}

\caption{Graphical representation of the  $1\to p$ QCD Pomeron amplitude. White
circles: internal $\ehk$ vertex functions entering the one-loop integral. Black circles: complex conjugates  external vertices $\bar \ehk$ coupling the external gluons to the interacting BFKL Pomerons. Double lines: BFKL Pomerons.}
\label{fig:vertices}
\end{figure}

\section{Phenomenology}
In the dipole model approach, hard diffraction in $\gamma^*$-proton processes is determined  \cite {bp} by the interaction between colour dipole states describing the photon and the proton.
Indeed, it is well-known that the photon can be analyzed in terms of $q\bar{q}$ 
configuration while
it has been shown \cite{mp} that the small-$x$ structure function of the proton
can be described by a collection of primordial dipoles with subsequent perturbative
QCD evolution. 
More specifically \cite{bp}, the combination of
the dipole description of perturbative QCD at high energy and the Good-Walker
mechanism \cite{gw} leads to a unified description of the proton total and
diffractive structure functions \cite{bpr}.

In the dipole approach, two components are shown to contribute to the diffractive
structure function. First, a {\it quasi-elastic}  component
 corresponds to the elastic interaction of two dipole configurations.
It is expected to be
dominant in the finite $\beta$ region, i.e. for small relative masses 
of the diffractive system. Interestingly,  it is related \cite {bianav} to the solution of the
conformal coupling of a $q\bar{q}$ state to the BFKL Pomeron.

Second, there is an {\it inelastic} component  where the initial
photon dipole configuration is diffractively dissociated in multi-dipole states
by the target. This process is expected to be important at
small $\beta$ (large masses). In this case, the theoretical calculation is directly related \cite {value} to the triple QCD Pomeron vertex, which is a particular case of the vertices considered in the previous section. Indeed, an explicit evaluation of the quantity $B_{1\to 2}$ (see equation (\ref {2})) has been performed using the dipole model  and leads to a rather large value \cite {value} of the triple QCD Pomeron coupling  corresponding to the calculation of $B_{1\to 2}$ at the Pomeron saddle-point values $h_0=h_1=h_2=\frac 12.$

Thus, in the QCD dipole model approach of hard diffraction, the conformal properties of the BFKL Pomerons through their couplings and vertices are 
relevant and could be tested by the phenomenological approach. The paper \cite {munier}  contains  a fit of the published diffractive data \cite{F2DH194}
 with 7 free parameters.  
It also includes a phenomenological secondary Regge trajectory 
which is known to play a r\^ole in the limited domain of large mass and
small rapidity gap \cite{F2DH194}.  
The fit is successful showing that this approach is a good candidate for a deeper 
understanding of hard diffraction. Note that non-leading logs perturbative corrections are phenomenologically taken into account via the parametrisation of an effective 
BFKL singularity with intercept lower than the bare value. Also, effective vertices of the {\it primordial dipole} distribution in the non perturbative proton target are not theoretically known but verify certain constraints, as discussed in detail in \cite {mp}. Both these aspects deserve more study in the future.
\section{Conclusion and outlook}
The dipole approach to QCD at high energy appears to be quite successful in analyzing deep inelastic scattering  at small  $x$ and in particular hard diffraction processes. At the theoretical level, it allows an evaluation of the complicated QCD multi-Pomeron vertices among which the triple Pomeron coupling is of relevance in the phenomenological studies. Recently also the first calculation \cite {navpesch} of 1-loop Pomeron contributions have been performed in the same framework.

Among the problems to be addressed to in the near future,  the intriguing  relationship of the QCD multi-pomeron vertices with conformal field theories arises the question whether the already known \cite{Lipatov} global conformal invariance of the BFKL kernel can be enlarged to some kind of Virasoro algebra. On the phenomenological ground,  the application of the QCD dipole model to   Tevatron results on diffraction and their comparison with HERA emerge as one of the most interesting questions.
\section{Acknowledgements}
I want to associate to this brief review A.Bialas, R.Janik, S.Munier, H.Navelet and Ch.Royon with whom the reported results have been obtained.
\eject

\end{document}